# 4-bit Factorization Circuit Composed of Multiplier Units with Superconducting Flux Qubits toward Quantum Annealing

**Daisuke Saida, Mutsuo Hidaka, and Yuki Yamanashi, *Member*, *IEEE***

This work is partly supported by the NEC-AIST Quantum Technology Cooperative Research Laboratory. This paper is partly based on results obtained from a project commissioned by the New Energy and Industrial Technology Development Organization (NEDO), Japan (project number JPNP16007).

(*Corresponding author: Daisuke Saida*)

D. Saida is with National Institute of Advanced Industrial Science and Technology (AIST), Tsukuba 305-8568, Japan. Current main affiliation is Fujitsu Laboratory, Fujitsu Limited, Kawasaki 211-8588, Japan (e-mail: saida.daisuke@aist.go.jp, saida.daisuke@fujitsu.com). M. Hidaka is with AIST, Tsukuba 305-8568, Japan. Y. Yamanashi is with Electrical and Computer Engineering, Yokohama National University, Yokohama 240-8501, Japan.

***Abstract*—Prime factorization ($P = M \times N$) is considered to be a promising application in quantum computations. We perform 4-bit factorization in experiments using a superconducting flux qubit toward quantum annealing. Our proposed method uses a superconducting quantum circuit implementing a multiplier Hamiltonian, which provides combinations of $M$ and $N$ as a factorization solution after quantum annealing when the integer $P$ is initially set. The circuit comprises multiple multiplier units combined with connection qubits. The key points are a native implementation of the multiplier Hamiltonian to the superconducting quantum circuit and its fabrication using a Nb multilayer process with a Josephson junction dedicated to the qubit. The 4-bit factorization circuit comprises 32 superconducting flux qubits. Our method has superior scalability because the Hamiltonian is implemented with fewer qubits than in conventional methods using a chimera graph architecture. We perform experiments at 10 mK to clarify the validity of interconnections of a multiplier unit using qubits. We demonstrate experiments at 4.2 K and simulations for the factorization of integers 4, 6, and 9.**

**INDEX TERMS** **Quantum annealing, prime factorization, superconducting flux qubit, multiplier**

# I. INTRODUCTION

For certain problems, quantum computers have the potential to drastically reduce computation times below what is required for classical computing. One of the most notable examples is prime factorization ($P = M \times N$). The number of steps a classical computer running currently known algorithms requires to find the prime factors of an $n$-digit integer $P$ increases exponentially with $n$. Therefore, factorization of large integers is classically difficult, and that is the basis for the security of widely used cryptographic codes. However, quantum computers can factor integers in polynomial time. The most famous prime factorization algorithm is Shor's algorithm [1], and there have been studies of its implementation in quantum circuits based on various physical phenomena [2–9]. Gate-type superconducting quantum circuits are currently gaining attention, due to demonstrations of the feasibility of implementing this algorithm in commercially available quantum computers. Key outstanding issues include clarifications of how to modify Shor's algorithm to implement noisy intermediate-scale quantum computers and their scalability.

Reference [10] reports the implementation of a simplified Shor's factorization algorithm in superconducting quantum circuits. However, there was a significant discrepancy between theoretical and experimental results for the prime factorization of 21. This was mainly due to cumulative error caused by the increased number of 2-qubit gates. This indicates that the number of bits must be increased with high fidelity. Another innovation was using quantum annealing (QA), in which prime factorization is treated as an optimization problem where solutions are the global minimum of the Hamiltonian [11]. However, this method requires classical computation to calculate the Gröbner basis, which helps to reduce the cost function. Deriving a cost function based on a multiplication table representing the integer $P$ has also been proposed as a QA method [12]. However, implementing the Hamiltonian into the chimera graph [13] requires many tasks, including the conversion of high-order terms to lower-order terms in the interactions. Generally, extra qubits must be introduced when constructing a new Hamiltonian. Specifically, the annealing dynamics can be changed as the spectra structure including the excited state is modified [14–15].

We demonstrate a scalable method that can perform prime factorization solely on a quantum computer. Our proposed method uses a superconducting quantum circuit implementing a multiplier Hamiltonian, which provides an $M$ and $N$ combination as a solution for a given $P$ after QA [16, 17]. As in a classical $n$-bit multiplier [18, 19], the circuit implements combinations of multiplier unit (MU) Hamiltonians [20]. Using this circuit, the combination of qubits (corresponding to $M$ and $N$) with the lowest energy configuration in the Hamiltonian appears as the solution after QA, depending on the initial condition expressing integer $P$. Figure 1(a) shows a schematic for an $n$-bit factorization chip. The chip includes multiple MUs, with $MU_{ij}$ ($i$ = 0-1, $j$ = 0-1) having $X_{ij}$ and $Y_{ij}$ as inputs, $Z_{ij}$ for sum-in, $D_{ij}$ for carry-in, $C_{ij}$ for carry-out, and $S_{ij}$ for summation. Using the ground state spin logic [21], $MU_{ij}$ can be expressed as a Hamiltonian. We have fabricated dedicated circuits for this MU Hamiltonian, which is implemented as superconducting flux qubits with all-to-all connectivity [16]. A 4-bit factorization circuit is formed using 4 MUs. Figure 1(b) shows the Hamiltonian of the 4-bit factorization circuit. Circles in that figure correspond to qubits, and the numbers within them show amplitudes of the local bias field. Lines between the circles and numbers indicate couplings between the qubits and their relative intensity. The formation of this circuit can be described as $P = (P_4\, P_3\, P_2\, P_1)_{(2)} = (C_{11}\, S_{11}\, S_{10}\, S_{00})_{(2)}$, highlighted in green in the figure. Inputs are represented as $M = (X_2\, X_1)_{(2)} = (X_{01}\, X_{00})_{(2)}$ and $N = (Y_2\, Y_1)_{(2)} = (Y_{11}\, Y_{01})_{(2)}$, respectively highlighted in blue and purple. Here, we utilize a binary number representation. In the Hamiltonian of the factorization circuit, we also insert an interconnection coupling between corresponding qubits where the carry propagations exist and the input terminals are identical. This method corresponds to expressing 1 equivalent qubit using 2 qubits belonging to different unit lattices [13, 22–27]. Note that the distance between 2 qubits belonging to different lattices is not constant when implementing the Hamiltonian shown in Fig. 1(b) in the superconducting quantum circuit. If a variable coupler is used for the lattice interconnection, 2 constraints on the coupler are required [28]. One constraint comes from the shape of an energy potential for the variable coupler: the value of a dimensionless factor $\beta_L$ ($\beta_L = 2\pi L I_c/\Phi_0$, where $L$, $I_c$, $\Phi_0$ are respectively the inductance, critical current, and flux quantum) should be less than 1.2 [28]. The other constraint is due to requirements for increasing the coupling intensity, from which the value of $L$ in the coupler must be less than 1/3 of the qubit inductance. However, even if these constraints are adhered to, the amplitude that the coupling can achieve is at most a few pH, making use of the variable coupler unsuitable for the interconnection. In this study, we investigate a connection qubit (CQ) for inter-grid routing. To suppress errors in the interconnections between qubits, we analyze the $L$ of the CQ through simulations. The 4-bit factorization circuit is fabricated using superconducting flux qubits. We evaluate features of the circuit in both simulations and experiments.

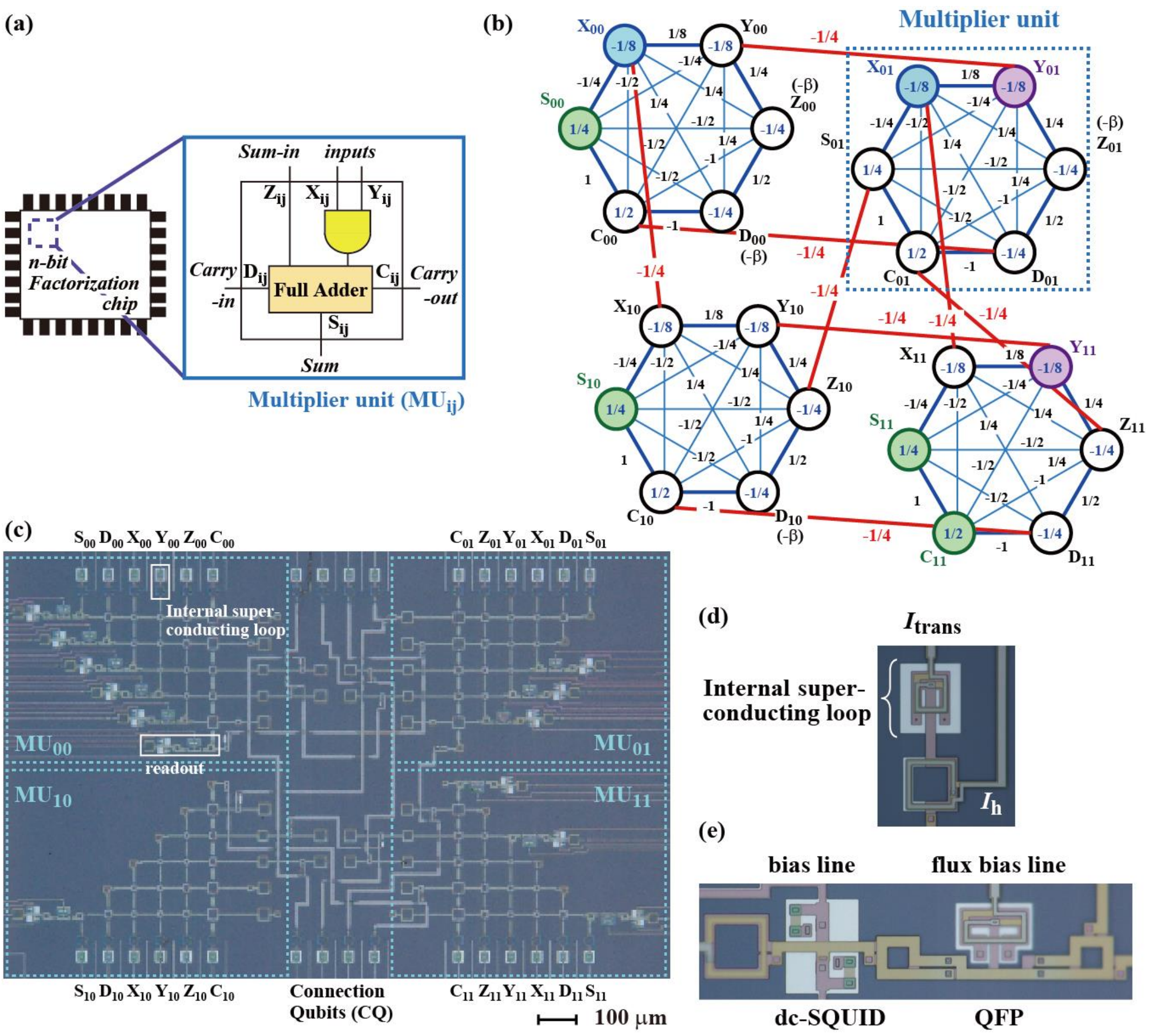

Fig. 1. (a) Schematic of the $n$-bit factorization chip based on multiplier units ($MU_{ij}$). (b) Description of Hamiltonian in the 4-bit factorization circuit. Production can be described as $P = (P_4\,P_3\,P_2\,P_1)_{(2)} = (C_{11}\,S_{11}\,S_{10}\,S_{00})_{(2)}$. Inputs are represented as $M = (X_2\,X_1)_{(2)} = (X_{01}\,X_{00})_{(2)}$ and $N = (Y_2\,Y_1)_{(2)} = (Y_{11}\,Y_{01})_{(2)}$, respectively. (c) Superconducting quantum circuit embedding 4-bit factorization Hamiltonian. The circuit is composed of 4 MUs and 8 connection qubits, consequently use of 32 superconducting flux qubits. Enlarged images of the area circled with white of (d) the internal superconducting loop and (e) the readout.

## II. QUBIT DESIGN FOR LATTICE INTERCONNECTION

We investigated the properties of CQs suitable for interconnections between MUs. When qubits belonging to different MUs are coupled by CQs, they prefer to take identical states after QA. Identical states in coupled qubits must be highly reproducible in each QA. Note that this depends on the inductance of the CQs. We use a Josephson integrated circuit simulator (JSIM) [29] to evaluate the error rate at which coupled qubits fail to be in identical states. We consider a model in which 2 qubits ($Q_1$, $Q_2$) belonging to different lattices are connected by the CQ. Figure 2(a) shows the model for the JSIM analysis. $I_h$, $I_{trans}$ and $I_p$ correspond to the current flowing through the local bias line, the current for the transverse field and the persistent current in the qubit, respectively. We simulate the error rate after QA with respect to the inductance value of CQ ($L_{CQ}$). To favor $Q_1$ adopting the 1 state, the flux is applied through the local bias current ($I_{h1}$). Calculations are performed under the same current condition of $I_{trans}$ in $Q_1$, CQ and $Q_2$. We distinguish success in the analysis whether $Q_1$ and $Q_2$ are in phase or not after the QA. We used 100 iterations for each condition. For convenience, the data is plotted as $10^{-2}$ if no error occurs during the iterations. As in our previous work [30], we use an annealing time ($T_a$) of 1 μs. We use 299.5 pH as the value for inductances of $Q_1$ and $Q_2$, this being the average inductance value for the qubits forming the MU. Figure 2(b) shows a simulation of the error rate with identical states between $Q_1$ and $Q_2$. The numbers of errors exceeds the limit of the number of trials above 400 pH. To avoid errors caused by routing with CQs, we designed the superconducting quantum circuit so that $L_{CQ}$s are less than 250 pH. A 4-bit factorization circuit requires 8 CQs. The mean and standard deviation for $L_{CQ}$ are 222 pH and ±5 pH, respectively (Fig. 2(c)). The area facing the qubits and CQ is designed to have a mutual inductance $M$ of 12 pH. This corresponds to 1/4 the magnitude in terms of the pairwise couplings $J_{ij}$ of the Hamiltonian in Fig. 1(b).

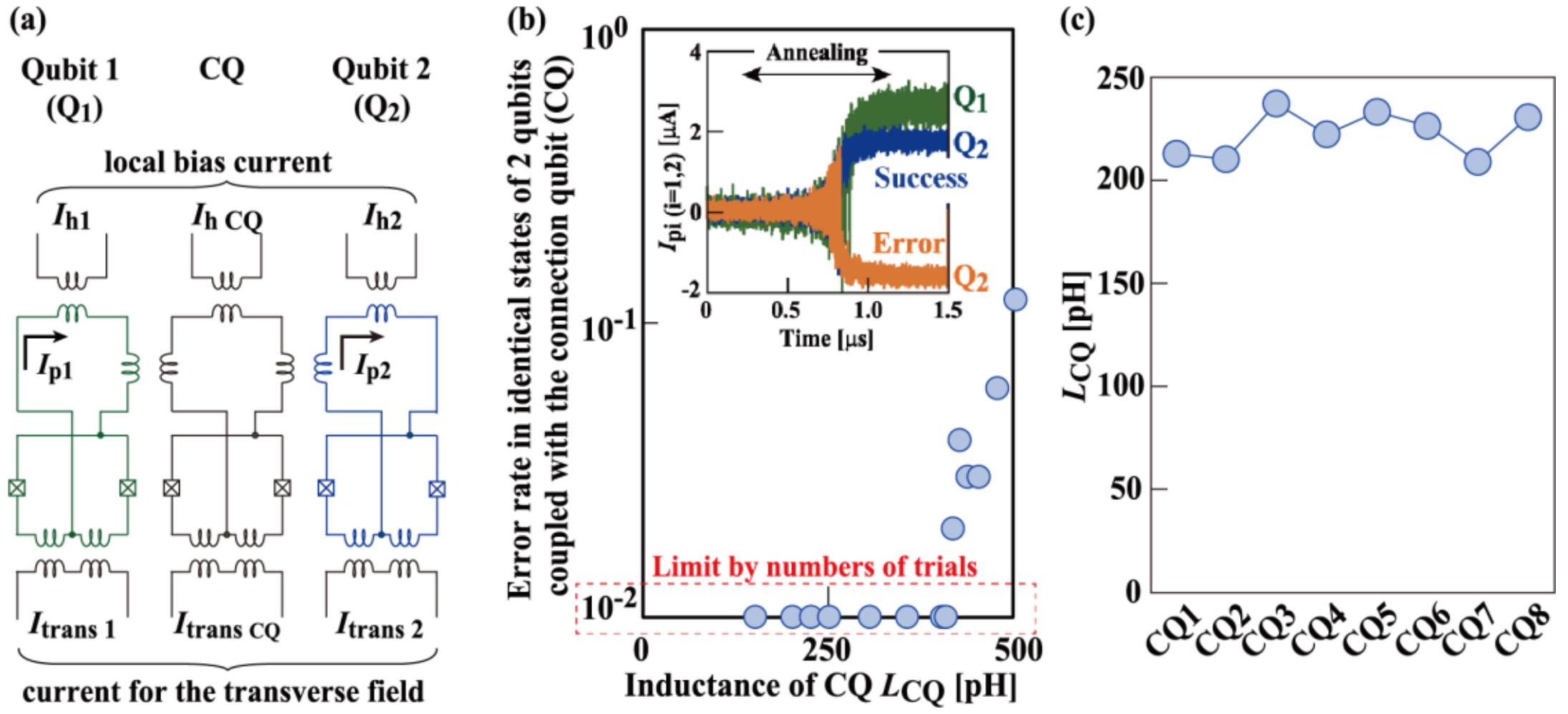

Fig. 2. (a) Circuit model expressing two qubits coupled with the connection qubit (CQ). (b) Simulation result of error rate in identical states of 2 qubits coupled with the CQ after annealing for 1 μs. (c) Designed inductances $L_{CQ}$ of 8 CQs implemented in the 4-bit factorization circuit. $L_{CQ}$s are designed within the range of 222 pH±5 pH.

## III. DEVICE FABRICATION AND EXPERIMENTAL SETUP

The qubits used in this experiment are superconducting compound Josephson junction rf-SQUID flux qubits, having a similar configuration to those in our previous works [16, 20, 30]. We fabricated the superconducting quantum circuit using a process that creates 4 Nb layers and a Josephson junction with a critical current density of 1 μA/μm$^2$ [31]. We adopted a 2.5 μm$^2$ Josephson junction for the superconducting flux qubit. Figure 1(c) shows an optical micrograph of the 4-bit factorization circuit composed of 4 MUs and 8 CQs. We implemented 32 superconducting flux qubits. Note that 16 unit lattices (including 128 superconducting flux qubits) are required to straightforwardly implement the Hamiltonian depicted in Fig. 1(b) (where qubits with all-to-all connectivity are utilized to express the MU Hamiltonian) to the conventional chimera graph. In effect, the Hamiltonian is implemented with 94 qubits. Our original method of fabricating the superconducting quantum circuit, which can directly implement the Hamiltonian, has the advantage of reducing the number of qubits. In QA, a local bias field $h_i$ and a transverse field $J_{ij}$ corresponding to the Hamiltonian coefficients are applied. Figure 1(d) shows enlarged image around the internal superconducting loop in the qubit. $I_h$ is the local bias current for $h_i$ and $I_{trans}$ is the current for the transverse field. The annealing schedule is controlled by the rise time in $I_{trans}$ [20]. The maximum amplitude of $I_{trans}$ corresponding to injection of the quantum flux $\Phi_0$ ($2.07 \times 10^{-15}$ Wb) applies to the internal superconducting loop in the qubit. The qubit state is detected through a readout circuit comprising a quantum flux parametron (QFP) and a dc superconducting quantum interference device (SQUID), as shown in Fig. 1(e). After QA, the qubit has 2 bistable states with persistent current flowing clockwise or counterclockwise. We define these 2 states so that they correspond to logical 1 and 0 states. Our experiments used a dilution refrigerator to cool the superconducting quantum circuit to 10 mK. We also conducted experiments at 4.2 K using a helium dewar.

## IV. FEATURE OF THE CONNECTION QUBITS

We evaluated the features of state transitions of 13 superconducting flux qubits in the 4-bit factorization circuit at 10-mK. We observed undesirable state transition features in the 4 qubits (see Supplementary Figs. 1(a), (b) and Appendix A). The readout circuit of each of the 4 qubits responded correctly against the applied magnetic field, indicating generation of flux trapping to the superconducting flux qubit, thus subjecting the large offset magnetic field in the state transition. In this state, the whole circuit cannot operate with the local bias current in the range of ±15 μA. We thus investigated proper operations of the CQ, which played a functionally important role in the 4-bit factorization circuit.

We used the QA to investigate the states of $Y_{00}$ and $Y_{01}$ qubits connected by the CQ. We use 1 to denote a qubit taking state 1 after QA, and 0 otherwise. Figure 3(a) shows a phase diagram expressing the states of $Y_{00}$ and $Y_{01}$ qubits with respect to local bias currents. Probabilities of each qubit state are evaluated experimentally under each bias condition with $10^4$ iterations at 10 mK. The shading represents the probability of each state from 1 to 0. In this experiment, the local bias field was not applied to the CQ. Regions of 11 (red) and 00 (blue) indicate that $Y_{00}$ and $Y_{01}$ qubits take identical states after QA. We found transition areas between each region. For convenience, we define the transient width between 2 regions as a “gray zone”. Widths of the gray zones are 0.10, 0.10, 0.15, and 0.18 μA between regions 10-11, 00-11, 10-00, and 11-01, respectively. The former values are 60–70% of the latter, suggesting they can easily change their energy state. We focus on the gray zone boundary between regions 10-00 and 11-01, where the local bias current $I_{h2}$ can form a wavy boundary. When $I_{h1}$ is fixed within the gray zone, oscillatory transition of the probability appears in the 11 and 00 states with respect to $I_{h2}$ (Figs. 3(b)(i), 3(b)(ii)). By contrast, the state transition probability, where $I_{h2}$ is fixed and $I_{h1}$ is modulated, has the features shown in Fig. 3(c). A closer examination reveals that no oscillations appear in the probability of “11,” but they do appear in the probability of “00” (see Supplementary Fig. 2 and Appendix B). These features—different gray-zone widths and wavy boundary shapes— may imply energy instability when only one of the qubits belonging to different lattices changes states by applying the local bias current. Our method first searches for a degeneracy point where all minimum energy states occur. Later, QA is performed by applying an offset current a to that point as an initial condition [16, 30]. To avoid undesirable effects in the gray zone, applying a large offset current may contribute to success in the experimental factorization. In the 11 region, we confirmed that coupled $Y_{00}$ and $Y_{11}$ qubits took identical states when the local bias current was solely applied to the 1-state in $Y_{00}$ (see Supplementary Fig. 3 and Appendix B), validating the functionality of the CQ for the inter-connection.

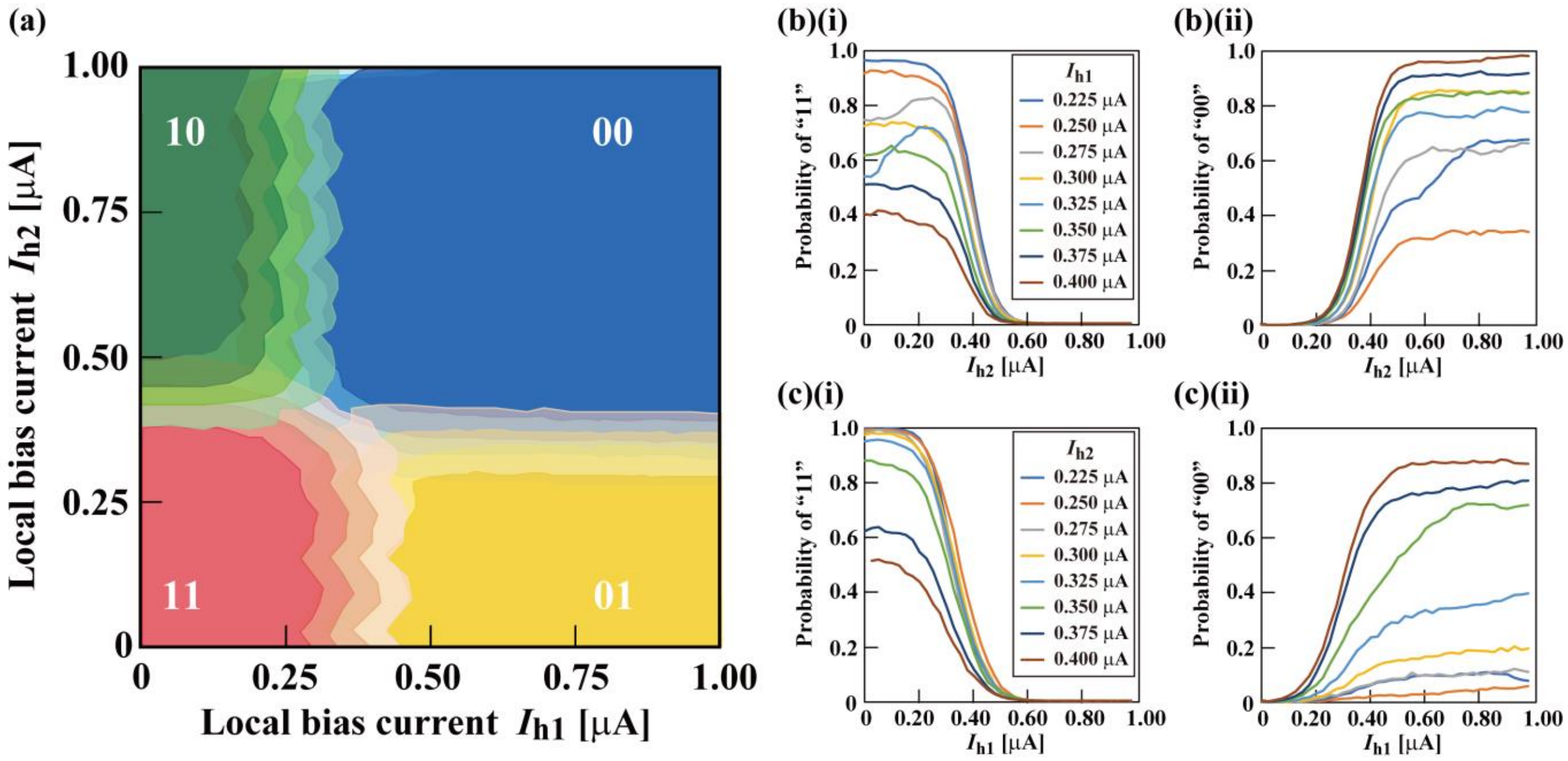

Fig. 3. (a) Phase diagram of expressing states of $Y_{00}$ and $Y_{01}$ qubits in the 4-bit factorization circuit with respect to the local bias currents at 10 mK. Boundaries of 10-00 and 11-01 are wavy along the sweep direction of $I_{h2}$. (b) Probability of (i) 11 states and (ii) 00 states with modulation of $I_{h2}$ in the fixed $I_{h1}$ condition. In both cases, oscillations appear in the state transition owing to the wavy boundary in the phase diagram. (c) Probability of (i) 11 states and (ii) 00 states with modulation of $I_{h1}$ in the fixed $I_{h2}$ condition. Experiments are carried out at 10 mK.

## V. FACTORIZATION

### *A. Experimental Demonstration at 4.2 K*

In the 10 mK experiment, the flux trap could not be released by repeatedly raising and lowering the temperature. We installed the 4-bit factorization chip in the helium dewar and cooled it to 4.2 K in order to repeatedly raise and lower the temperature for defluxing while checking the wiring conditions in the chip with those resistances each time. The energy in the potential of the qubit around its bottom is about $2.0 \times 10^{-21}$ J. The thermal energy at 4.2 K is estimated to be $5.8 \times 10^{-23}$ J. At this energy, thermal noise affects the qubit, so it is no longer quantum annealing. Therefore, we distinguish the experiment at 4.2 K from QA. However, it is possible to check whether the circuit is responsive based on the Hamiltonian.

One qubit took a feature affected by the offset magnetic field in the state transition evaluation (see Supplementary Fig. 1(c)). This indicates the generation of flux trapping. Fortunately, the whole circuit can be operated with the local bias current between –5 and 15 μA. Since the degeneracy point could not be theoretically predicted due to the existence of the offset magnetic field, we searched for current conditions occurring as combinations of qubit states having all the minimum energy states in the Hamiltonian of $MU_{00}$. Supplementary Fig. 4 shows a histogram of the $MU_{00}$ at currents ($I_{h1}$, $I_{h2}$, $I_{h3}$, $I_{h4}$, $I_{h5}$, $I_{h6}$) = (5.0, 4.2, 7.0, 3.5, 3.4, 3.0) μA over 10,000 iterations. All minimum energy states in the Hamiltonian are observed. Using this current as the degeneracy point, the offset current α is considered as an initial condition depending on the number to be factorized. For example, in the case of factorizing $P = 4 = (0100)_2$, annealing with $T_a$ of 100 μs is performed with local bias currents of

$$\left(I_{h_MU11Q5}{}', I_{h_MU11Q6}{}', I_{h_MU10Q6}{}', I_{h_MU00Q6}{}'\right) = \left(I_{h_MU11Q5} - \alpha, I_{h_MU11Q6} + \alpha, I_{h_MU10Q6} - \alpha, I_{h_MU00Q6} - \alpha,\right)$$

as the initial condition. The offset current a is modulated in the range 1–6 μA. By considering the Hamiltonian shown in Fig. 1(b), β of 7 μA is simultaneously considered as the initial condition. Figures 4(a)(i), (b)(i), and (c)(i) respectively show success probabilities for the factorizations of $4 = (0100)_2$, $6 = (0110)_2$, and $9 = (1001)_2$ with 10,000 iterations. Respective (ii) figures show the voltage signal in the dc-SQUID when the qubits corresponding to $X_{ij}$ and $Y_{ij}$ take successful factors. Correct factors of 2 × 3 and 3 × 2 were obtained for the factorization of 6. Though the success probability is low, we confirmed appropriate factors for all numbers that can be handled by the circuit. The low correct response rate is likely because the system is in a state under flux trapping with the existence of a large offset magnetic field. Therefore, increasing the offset current α did not improve the solution accuracy as intended. We also considered the value of β at the initial conditions, but this did not contribute to the intended modulation of the success probability (see Supplementary Fig. 5 and Appendix C). An outstanding task is thus to reform the qubit structure so that flux trapping does not occur.

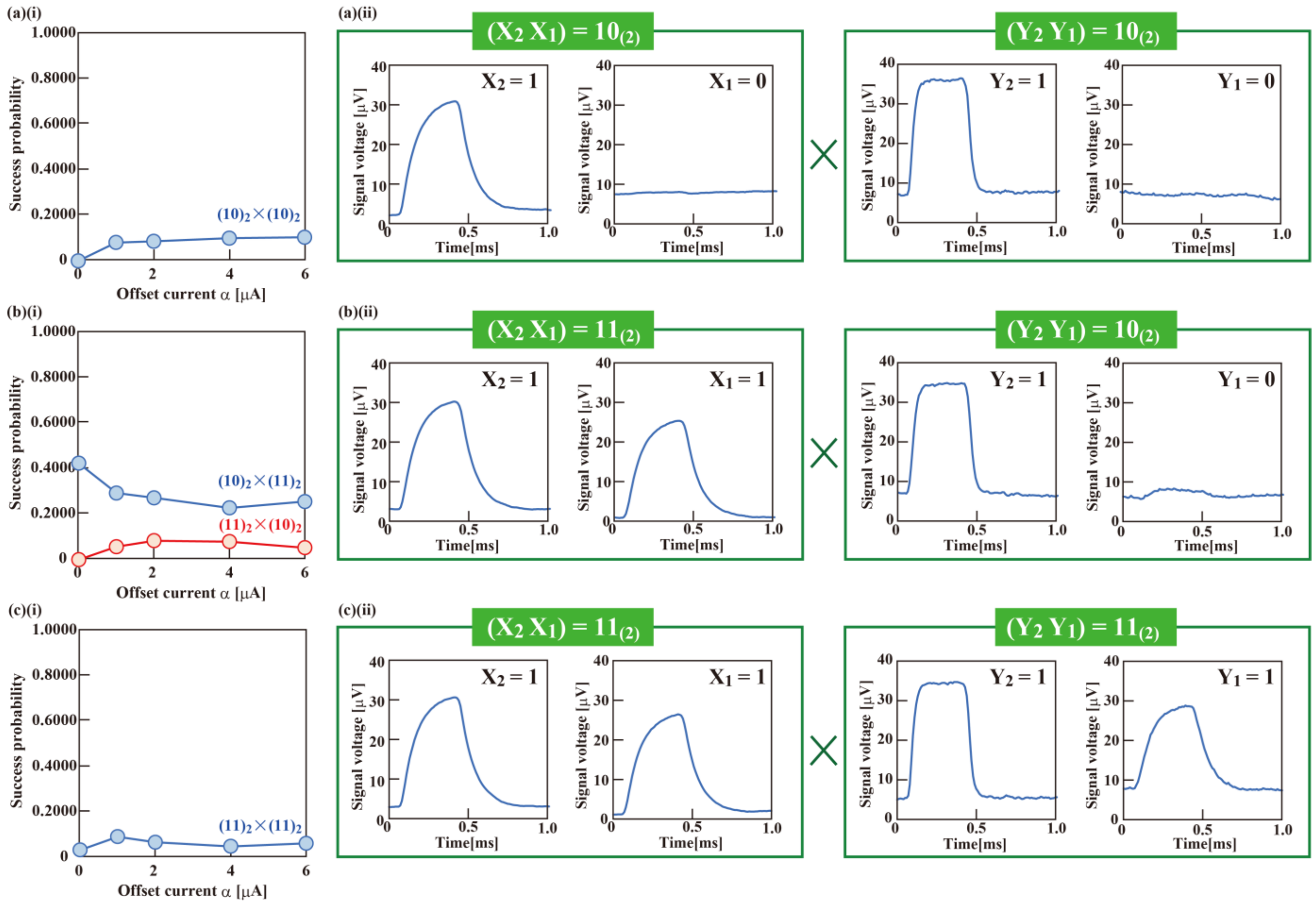

Fig. 4. (i) Success probability in factorization of integer 4 and (ii) voltage signals in the readout expressing input combinations of $M = (X_2\ X_1)_{(2)}$ and $N = (Y_2\ Y_1)_{(2)}$. (b) (i) Success probability in factorization of integer 6 and (ii) voltage signals expressing the appropriate combination of factors. (c) (i) Success probability in factorization of integer 9 and (ii) voltage signals. Experiments are carried out at 4.2 K.

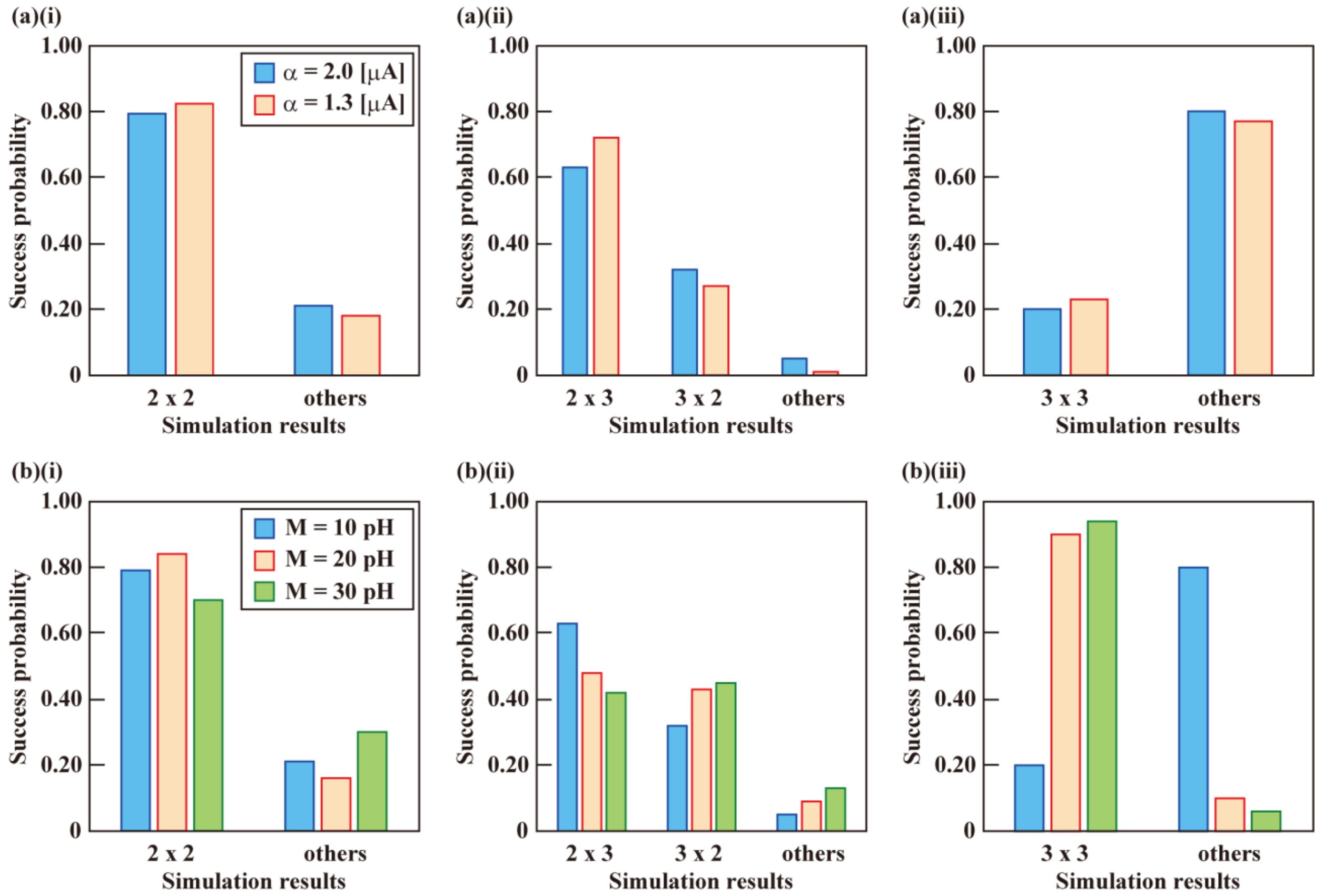

Fig. 5. (a) Success probabilities of factorizations of integers (i) 4, (ii) 6, and (iii) 9 in JSIM analysis. Two kinds of offset current α applied to the degeneracy point are considered. The mutual inductance *M* between the qubit and the CQ is 10 pH. (b) Success probabilities of factorizations of integers (i) 4, (ii) 6, and (iii) 9 at a of 1.3 μA with respect to modulation of *M* among10-30 pH.

### *B. Analysis of Success Probability in Factorization by Simulation*

Expressing the 4-bit factorization circuit shown in Fig. 1(c) as an integrated circuit emphasis model, we used JSIM to analyze the circuit characteristics. Supplementary Fig. 6 shows transient readout signals for integers 4, 6, and 9 at the initial conditions with $T_a$ of 1 μs (see Appendix D). Voltage signals of $V(X_{01})$ and $V(X_{00})$ and of $V(Y_{11})$ and $V(Y_{01})$ correspond to the $M = (X_2\,X_1)_{(2)} = (X_{01}\,X_{00})_{(2)}$ and $N = (Y_2\,Y_1)_{(2)} = (Y_{11}\,Y_{01})_{(2)}$, respectively. The existence of voltage (state-1) or its non-existence (state-0) at 1.07 μs in each readout signal indicates whether an appropriate factorization occurred.

Figures 5(a)(i), (a)(ii), and (a)(iii) respectively show success probabilities for factorizations of integers 4, 6, and 9. Here, we use mutual inductance *M* of 10 pH for interactions between the qubit and CQ. The offset currents a of 1.3 μA and 2.0 μA are applied. The success probability increases when α is 1.3 μA. Here, the MU, which includes the 4-bit factorization circuit, tends to take the highest success probability when α is 1.3 μA [16]. This feature may appear in the 4-bit factorization circuit. The simulation results indicate that success probabilities above 80% are possible for the factorizations of 4 and 6 if the flux trapping does not occur. By contrast, low success probabilities are indicated for the factorization of 9.

We performed JSIM analysis with an α of 1.3 μA, considering the modulation of *M* at 10–30 pH. Figures 5(b)(i), (b)(ii), and (b)(iii) respectively show success probabilities for factorizations of 4, 6, and 9. As a trend, we obtained high success probabilities with an *M* of 20 pH, and we can expect success probabilities above 80% for every integer. This indicates that there is not enough mutual inductance between the qubit and the CQ in the present circuit. In this circuit, we select an inter-lattice coupling *J* with a relative magnitude of 1/4, corresponding to an *M* of 12 pH. The simulation results indicated that inter-lattice coupling with a relative magnitude above 1/4 is suitable for correct operation. We will design our next circuit so that an $M$ of 20 pH is preferable for interactions between the qubit and the CQ.

### C. Guidelines and Discussion for Improvement of Device Features

As Supplementary Fig. 7 shows, the present superconducting flux qubit has unnecessary large structure in the M1 metal layer, which is not required for operation (see Appendix E). Although the magnetic shield reduces the strength of the magnetic field, the large area of the superconducting film may cause magnetic flux trapping, due to gathering flux exceeding the flux quantum during the cooling process. Therefore, a candidate design modification is to eliminate unnecessary superconducting film area, which may allow us to investigate the dependence of the offset current on the probability of a successful factorization and to perform QA in an environment with low thermal noise at 10 mK.

The JSIM analysis revealed that our current design lacks coupling strength between the MUs. This can be a factor behind errors resulting from inter-lattice connections by CQ. Therefore, we believe that a higher success probability can be expected by selecting an $M$ of 20 pH in the next qubit design.

## VI. CONCLUSION

We fabricated a 4-bit factorization circuit composed of superconducting flux qubits and extended by assembling multiple MUs as in a classical multiplier. We utilized CQs to interconnect the lattice MUs. The $n$-bit factorization chip comprises MUs and CQs, where $6(n/2)^2 + n^2 - 2n$ superconducting flux qubits are implemented. The 4-bit factorization circuit comprises 32 qubits. We confirmed that CQ worked as intended in experiments using a 4-bit factorization circuit at 10 mK. We found flux trapping to be a problem, probably due to an unnecessary structure at the internal superconducting loop in the qubit. We obtained correct solutions for integer factorizations in 4.2 K experiments, despite the magnetic flux trapping. This indicates circuit configuration is correct to express the Hamiltonian. We believe that revising the structure of the internal superconducting loop in the qubit will allow us to factorize the experiment at 10 mK. JSIM analysis indicated the possibility of factorization success probabilities above 80% by adopting an $M$ of 20 pH between the qubit and the CQ for lattice inter-connections. In future works, modifying the qubit structure and the CQ coupling strength should improve correct response rates and verify the effect of the QA at 10 mK. We believe our method, namely native implementation of the Hamiltonian to a superconducting quantum circuit with QA, can potentially overcome the scaling problem, opening the way for prime factorizations of large integers.

## APPENDIX A

### *Evaluation of The State-1 Probability in The Superconducting Flux Qubit*

In the first step, as a pretest of the 4-bit factorization circuit, we use the helium dewar to evaluate the probability of state transition of an individual qubit at 4.2 K. The selected appropriate chip, embedding the 4-bit factorization circuit, is installed in the dilution refrigerator. In the circuit used in this study, pretesting indicated no qubit anomalies related to its state transition probability. Supplementary Fig. 1(a) shows the state-1 probability for an individual qubit at 10 mK. The behavior of the designed state transition can be found near the origin of $I_h$. Unintended behaviors appear in 4 qubits ($MU_{00}Q_1$, $MU_{00}Q_5$, $MU_{00}Q_6$, and $MU_{10}Q_6$). A state transition of the qubit $MU_{00}Q_5$ indicates the existence of an unintended offset magnetic field effect. The state transitions of qubits $MU_{00}Q_1$, $MU_{00}Q_6$, and $MU_{10}Q_6$ indicate a lack of control in the shape of the energy potential by the transverse magnetic field [33].

Supplementary Fig. 1(b) shows the state-1 transition when the 4-bit factorization circuit is heated to a temperature above that required for superconducting transition and then re-cooled to 10 mK. The behaviors of the 4 qubits exhibiting unintended transition properties are still confirmed. In the 10 mK experiment, the entire circuit could not be worked by a local bias current $I_h$ within the range ±15 μA. The 4 qubits produced the intended state transition when the pretest was performed at 4.2 K, suggesting that the offset field prevented the intended transition. The most likely reason is the generation of flux trapping.

We reinserted the 4-bit factorization circuit into the helium dewar and re-evaluated the state-1 probability for individual qubits at 4.2 K (Supplementary Fig. 1(c)). The 4 qubits ($MU_{00}Q_1$, $MU_{00}Q_5$, $MU_{00}Q_6$, $MU_{10}Q_6$) show the intended state transition, but qubit $MU_{00}Q_3$ showed an unintended state transition with the offset magnetic field. Though flux trapping occurs, the entire circuit could be worked by a local bias current $I_h$ within the range –5 to 15 μA.

## APPENDIX B

### *Detailed Evaluation of The Inter-Lattice Connection*

This section describes the phase diagram of 2 qubits coupled by CQs belonging to different unit lattices with respect to the local bias current (Chapter VI). A gray zone, corresponding to the transition region, is observed between the 4 states (00, 01, 11, 10). In particular, we focus on the wavy gray zone boundary between the 10-00 and 11-01 regions with respect to the $I_{h2}$. Supplementary Fig. 2 shows the probability of a coupled qubit state when $I_{h1}$ is modulated with a fixed $I_{h2}$ (within the gray zone). Oscillatory behavior does not appear in the probability of the 11 state (Supplementary Fig. 2(a)) and appears in the probability of the 00 state (Supplementary Fig. 2(b)). The change from the 11 state to the other states by modulation of $I_{h1}$ does not vary much with the $I_{h2}$ value (Supplementary Fig. 2(a)). On the other hand, features of the transition to the 00 state differ with the $I_{h2}$ value (Supplementary Fig. 2(b)). These features—different gray zone widths and a wavy boundary—may imply an energy instability when applying the local bias current changes the state of only one of the qubits belonging to different lattices. To avoid undesirable effects in the gray zone, we carefully consider applying an appropriate offset current to the degeneracy point.

We simultaneously measured the states of qubits $Y_{00}$ and $Y_{01}$ on an oscilloscope by applying the local bias current solely to the qubit $Y_{00}$ to produce the 1-state (Supplementary Fig. 3). Two qubits took identical states after the QA. In this experiment, the local bias current was not applied to the CQ. We expect calibration of the offset magnetic field from the surrounding circuit to increase the reproducibility of identical states in 2 qubits coupled by the CQ [34].

## Appendix C

### *Factorization Utilizing The Superconducting Quantum Circuit*

Our method first searches for the degeneracy point where the ground state of the Hamiltonian appears. The degeneracy point is predictable in theory [30], and calibrating the offset magnetic field affecting the surrounding circuit should allow predictions of the degeneracy point in experiments. However, most of the Hamiltonian ground states did not appear at the predicted degeneracy point because of the existence of an unintended large offset magnetic field, probably due to flux trapping. We thus searched for local bias current conditions within the range –5 to 15 μA in which all ground states appeared (Supplementary Fig. 4). The condition was that ($I_{h1}$, $I_{h2}$, $I_{h3}$, $I_{h4}$, $I_{h5}$, $I_{h6}$) = (5.0, 4.2, 7.0, 3.5, 3.4, 3.0) μA. Using this current as a reference, the offset current α is considered as an initial condition, one depending on the integer $P$ for which factorization is to be performed. In the initial condition, we simultaneously consider the β described in the Hamiltonian of Fig. 1(b). We considered 3 ways of doing so. The first is the method discussed in Chapter V, where β is given as an offset current greater than α. The second method is where β = α. Supplementary Figs. 5(a)(i), (b)(i), and (c)(i) show factorizations of integers 4, 6, and 9 in a 4.2-K experiment implementing the second method. In the third method, β is an offset current greater than α, and β is additionally applied for the qubit $MU_{01}Q_4$. Since the 4-bit factorization circuit does not carry digits from $MU_{00}$ to $MU_{01}$, intentional manipulation of the qubit state in $MU_{01}Q_4$ seems more likely to take zeros. Supplementary Figs. 5(a)(ii), (b)(ii), and (c)(ii) show factorizations of integers 4, 6, and 9 based on the second method in a 4.2-K experiment. However, in both the second and third methods, we observed no differences depending on the initial conditions because of the existence of a large offset magnetic field. This is probably due to the effects of magnetic flux trapping. It is thus necessary to modify the qubit structure so that magnetic flux trapping does not occur. To that end, we believe that removing unnecessary areas in the qubit is effective.

## APPENDIX D

### *Factorization Utilizing The JSIM*

We created the circuit model for the 4-bit factorization circuit as in Ref. [20], adopting experimental parameters for $L$ and $M$. Supplementary Fig. 6 shows the JSIM analysis in the factorization of integers 4, 6, and 9 with $T_a$ of 1 μs. The 4-qubit states, corresponding to the inputs $M = (X_2\ X_1)_{(2)} = (X_{01}\ X_{00})_{(2)}$ and $N = (Y_2\ Y_1)_{(2)} = (Y_{11}\ Y_{01})_{(2)}$, are expressed as the voltage signals observed in the readout circuit. Whether voltage exists (state 1) or not (state 0) at 1.07 μs in each readout signal indicates the appropriate factorization for each integer.

## APPENDIX E

### *Candidate Reasons for Flux Trap Generation*

Supplementary Fig. 7 shows a magnified optical micrograph of the area where magnetic flux is applied in the superconducting flux qubit. The transverse magnetic field and the local bias field are respectively applied through the currents $I_{trans}$ and $I_h$. The qubit has the SQUID structure, in which the qubit forms a main loop with M1- and M2-metal layers, which are shunted through Josephson junctions. In the present qubit design, the internal superconducting loop is constructed on the M1-metal rectangular area with a 30 μm width. This size is unnecessary when forming the main loop of the qubit. While the magnetic shield reduces the strength of the magnetic field in the dilution refrigerator, the large area of the superconducting film may cause magnetic flux trapping, due to gathering flux exceeding the flux quantum during the cooling process.

## ACKNOWLEDGMENT

We thank K. Inomata (AIST) and T. Kamimura (NF Corporation) for their technical support. We also thank F. Hirayama (AIST) for useful discussions. The devices used were fabricated in Japan at CRAVITY, AIST.

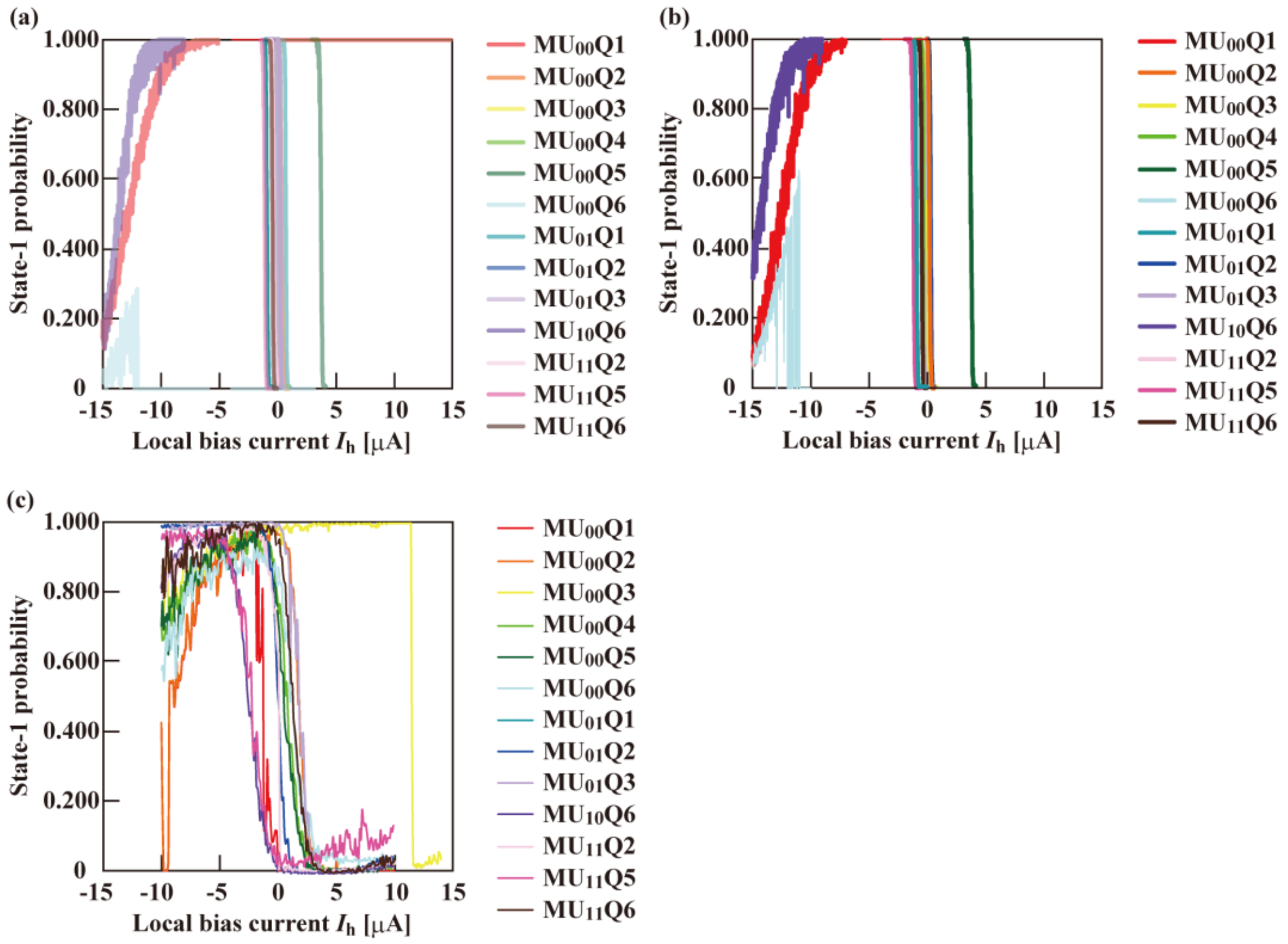

Supplementary Fig. 1. (a) State-1 probabilities of 13 qubits in the 4-bit factorization circuit evaluated at 10 mK. (b) State-1 probabilities when heated to the normal conduction state and re-cooled to 10 mK. Four qubits ($MU_{00}Q_1$, $MU_{00}Q_5$, $MU_{00}Q_6$, $MU_{10}Q_6$) indicate undesirable state transitions. (c) State-1 probabilities evaluated at 4.2 K. The state transition of qubit $MU_{00}Q_3$ indicates an influence of the offset magnetic field.

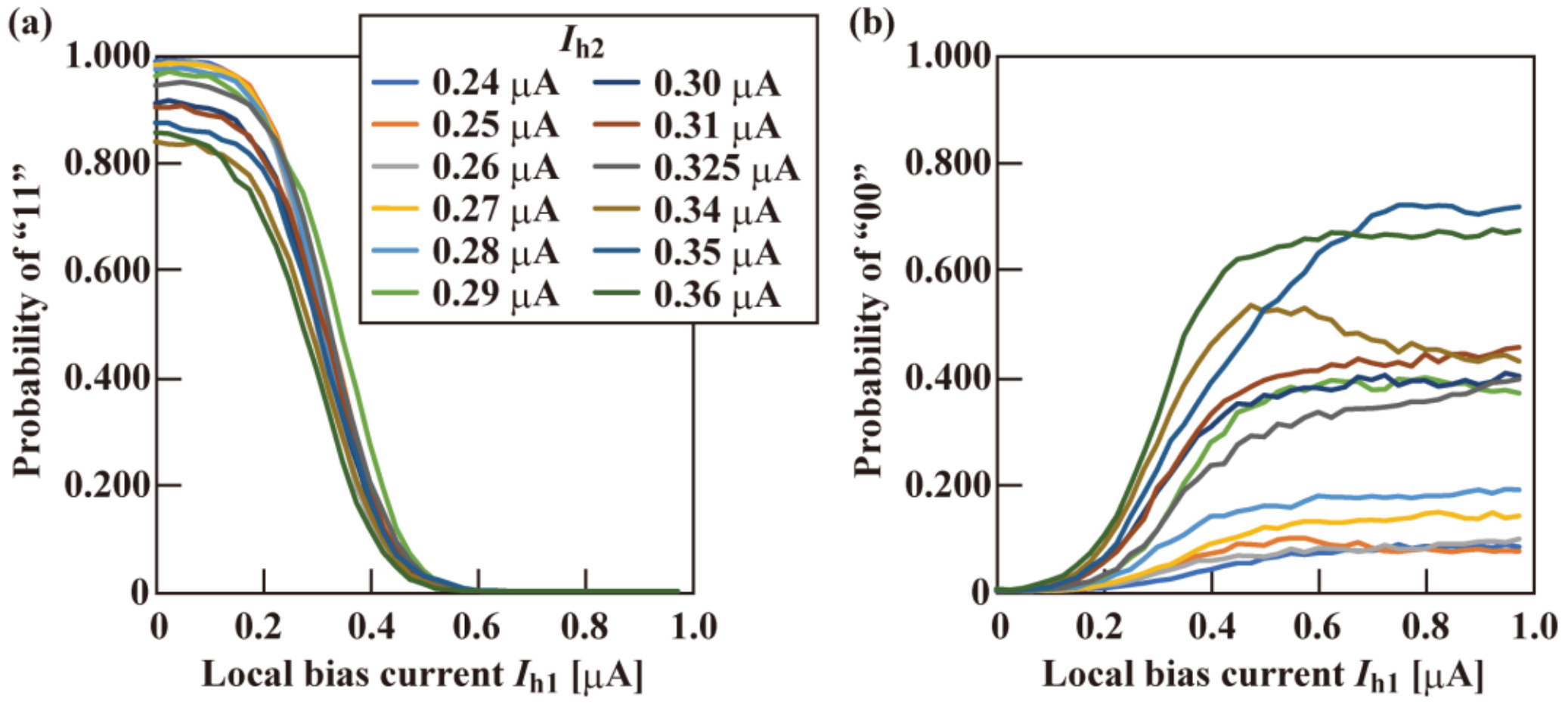

Supplementary Fig. 2. (a) Probability of the 11 state in a coupled qubit with modulation of local bias current $I_{h1}$ at the fixed condition of $I_{h2}$. This is a more detailed study of the conditions of the experiment shown in Figure 3(c)(i). No oscillatory behavior appears. Moreover, the state transition trends are similar within the range of measured currents. (b) Probability of the 00 state in a coupled qubit with modulation of local bias current $I_{h1}$ at fixed condition of $I_{h2}$. This is a more detailed study of the experimental conditions shown in Fig. 3(c)(ii). An oscillatory state transition is observed.

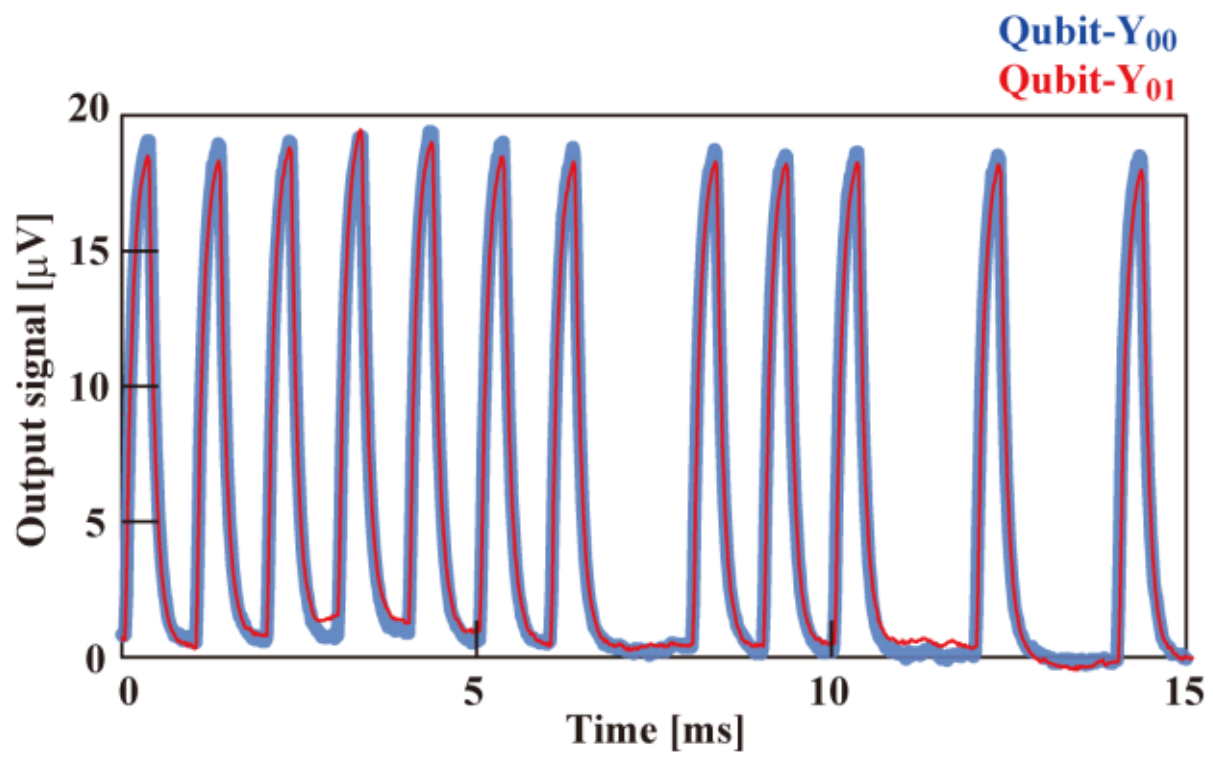

Supplementary Fig. 3. Output voltage signals in the readout detecting the states of qubits $Y_{00}$ and $Y_{01}$ in the 4-bit factorization circuit in the 10-mK experiment. Two qubits coupled by the CQ take identical states after QA. This result shows correct functionality of the CQ for the lattice inter-connection.

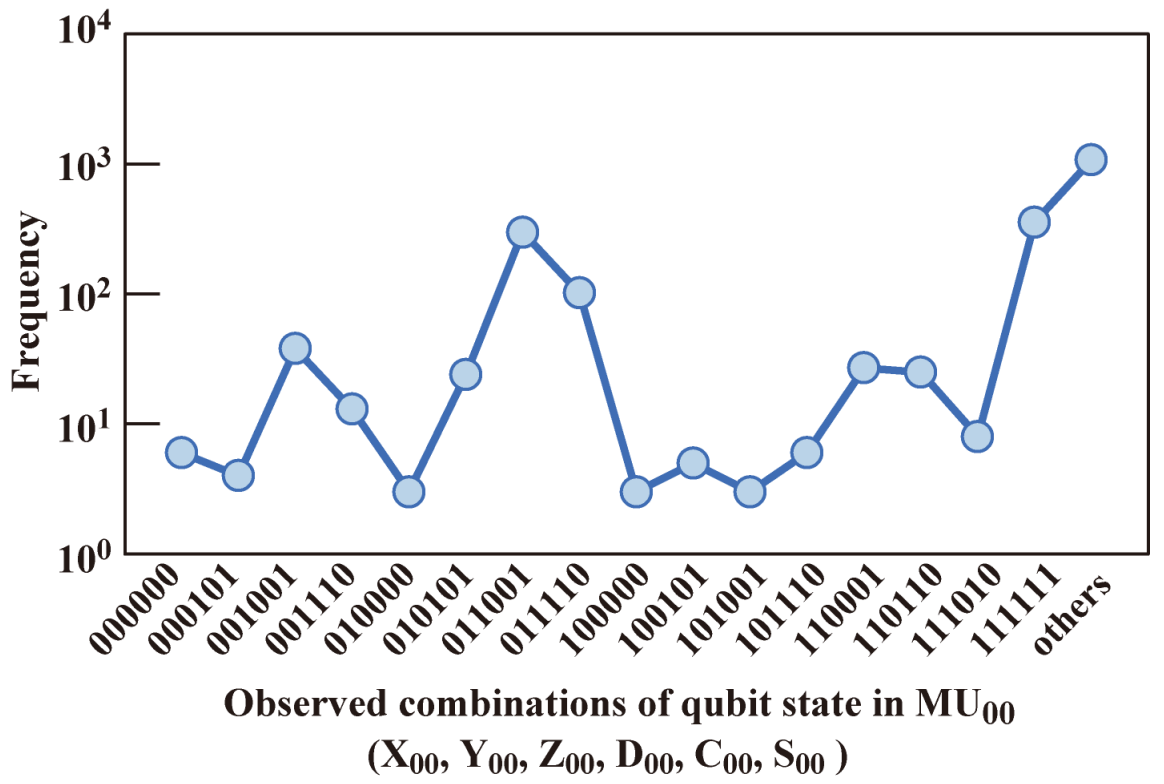

Supplementary Fig. 4. Frequency distribution of each logic component in the experiments at $10^4$ iterations using $MU_{00}$ in the 4-bit factorization circuit. $T_a$ is 100 μs. This experiment was carried out at 4.2 K. All 16 components corresponding to the Hamiltonian ground states are observed.

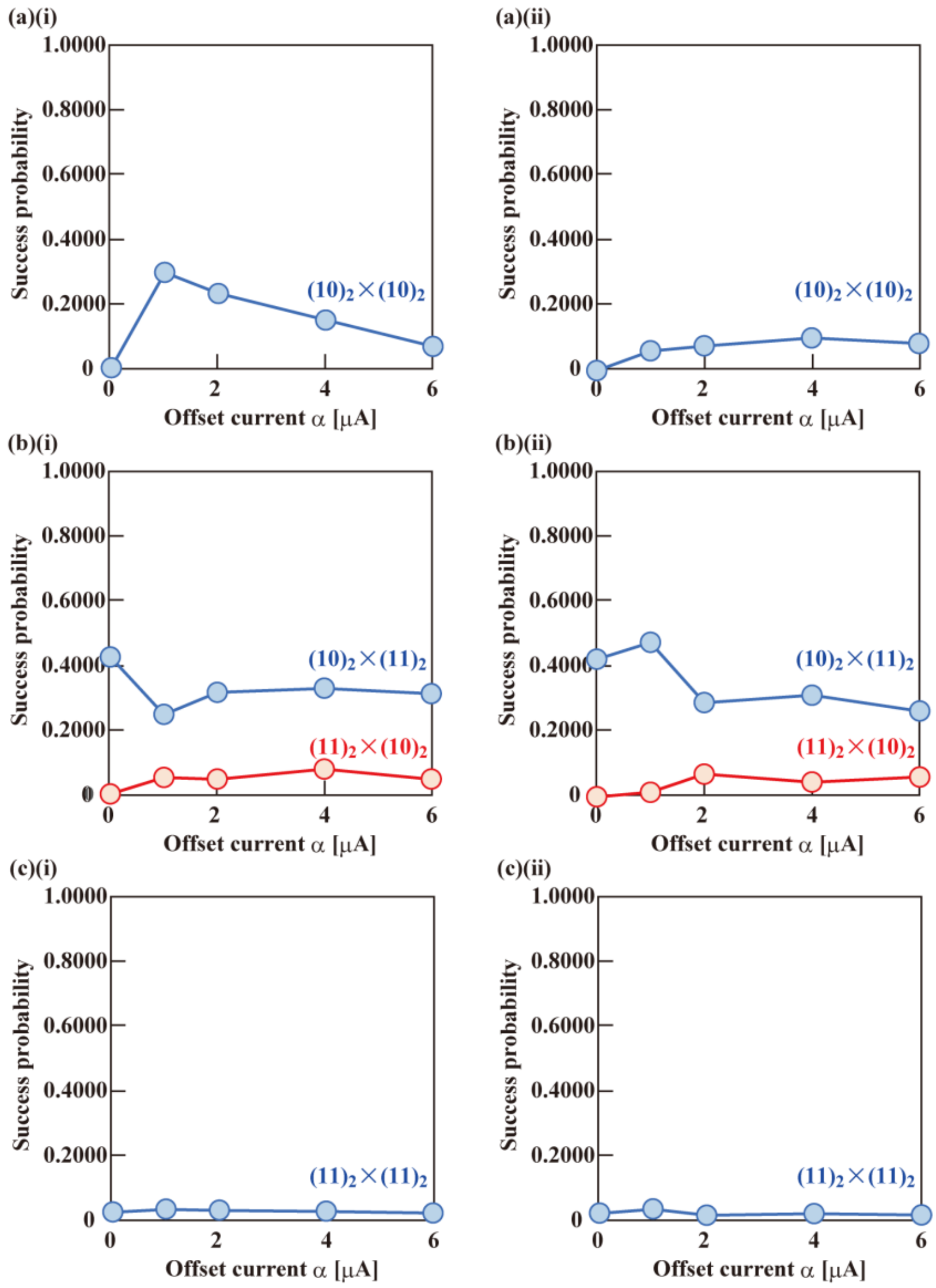

Supplementary Fig. 5. Success probability in the factorization of integers (i) 4, (ii) 6, and (iii) 9 with offset condition $\beta = \alpha$. (b) Success probability in the factorization of integers of (i) 4, (ii) 6, and (iii) 9 with offset condition $\beta > \alpha$ and addition of $\beta$ to the qubit $MU_{01}Q_4$.

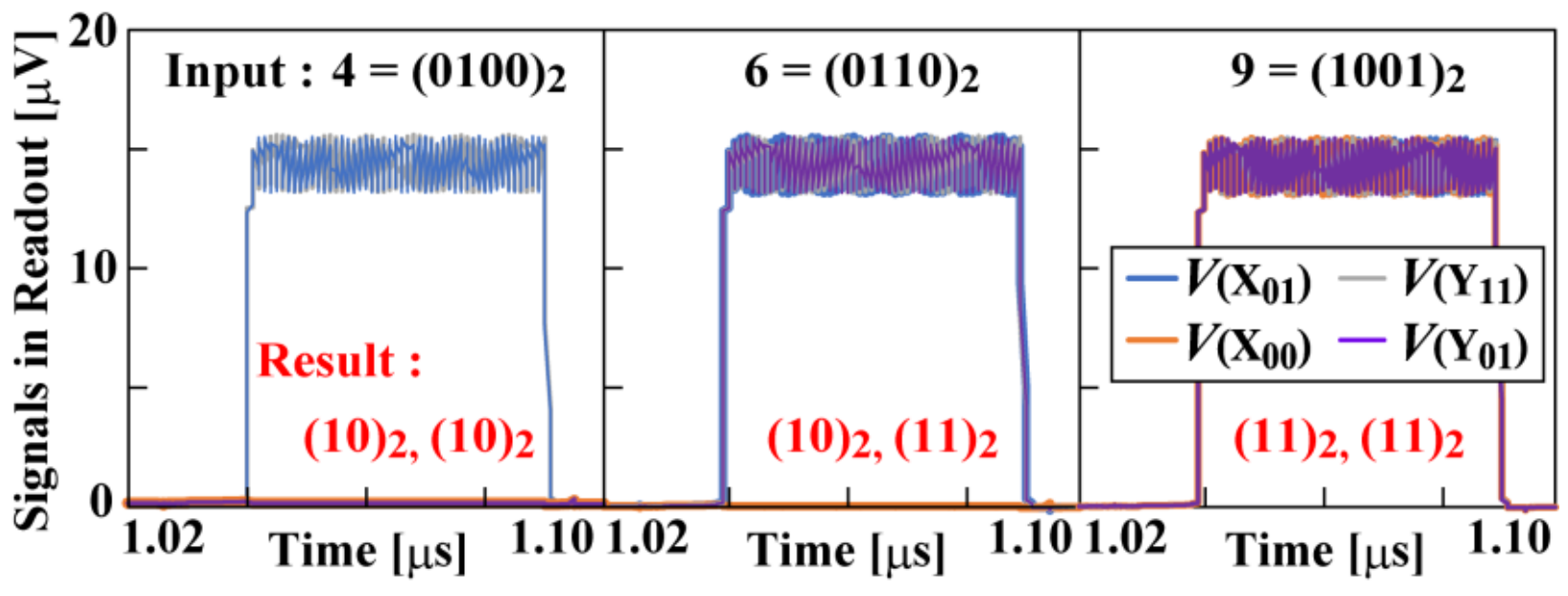

Supplementary Fig. 6. JSIM analyses in factorizations of integers 4, 6, and 9 with $T_a$ of 1 μs. After annealing, we can obtain ground states of the 4-bit factorization Hamiltonian as factors $M = (X_2\ X_1)_{(2)} = (X_{01}\ X_{00})_{(2)}$ and $N = (Y_2\ Y_1)_{(2)} = (Y_{11}\ Y_{01})_{(2)}$.

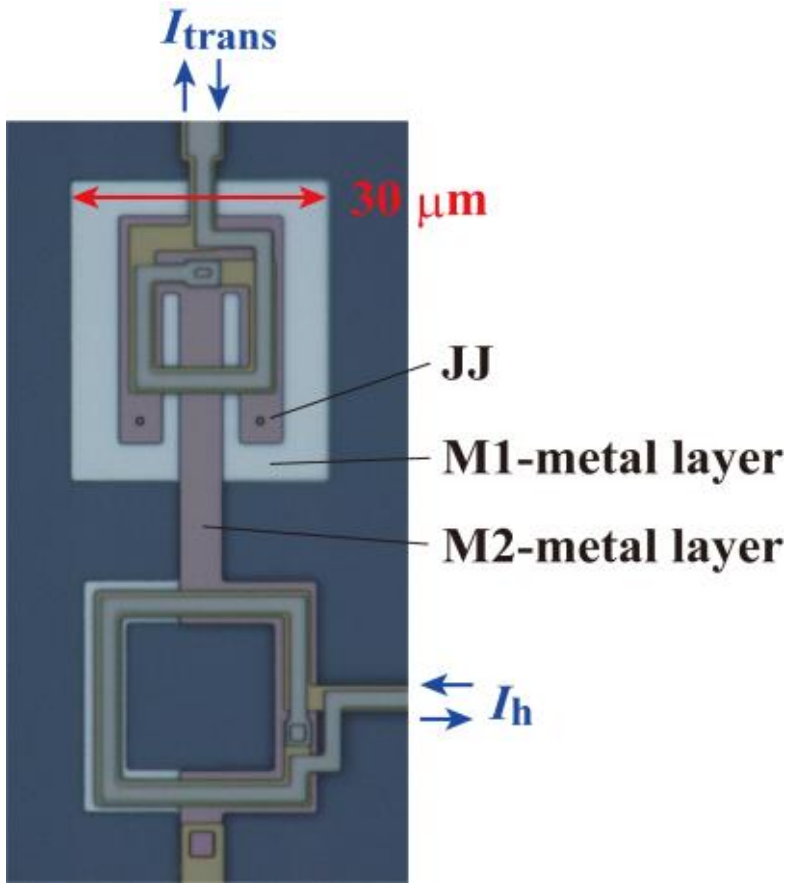

Supplementary Fig. 7. A magnified optical micrograph of the area where the magnetic field is applied in the superconducting flux qubit. A transverse magnetic field and a local bias field are respectively applied through the currents $I_{\mathrm{trans}}$ and $I_{\mathrm{h}}$. The qubit has the internal superconducting loop where unnecessary areas exist. This area probably causes the magnetic flux trapping.